\documentstyle[]{article}

\author{Ludger Hannibal}
\title{Geometric Foundation of Spin and Isospin
}
\author{Fachbereich Physik, Carl v. Ossietzky Universit\"at\\
 D-26111 Oldenburg, Germany\\
e-mail: hannibal@caesar.physik.uni-oldenburg.de}

\begin{document}

\maketitle
\begin{abstract}
Various theories of spinning particles are interpreted as realizing elements
of an underlying geometric theory. Classical particles are described by
trajectories on the Poincar\'e group. Upon quantization an
eleven-dimensional Kaluza-Klein type theory is obtained which incorporates
spin and isospin in a local $SL(2,C)\times U(1)\times SU(2)$ theory with
broken $U(1)\times SU(2)$ part.
\end{abstract}

The theories that describe the spinning property of electrons and other
particles developed completely different regarding the classical and the
quantum description. The classical approach initiated by Frenkel \cite
{Frenkel} and Thomas \cite{Thomas} was continued with many efforts \cite
{many,BTM,Nyborg}, but never
settled to a generally accepted theory, as recent work shows \cite
{Mukunda,BarutZanghi,BarutCo,vanHolten,YeeBander}. One apparent reason was
that its relation to the quantum theory, as developed by Dirac \cite{Dirac1}%
, never was fully clarified \cite{Nyborg,Ellis}. On the other hand,
 Dirac's theory
was soon accepted, supported by experiment, further developed and immersed
in the general framework of the representation theory of the Poincar\'e
group \cite{Wigner,representation}; 
today it forms the basis of particle description in
modern elementary particle physics.

There are two characteristic problems common to most classical theories of
spinning particles. The first is, that equations for momentum-like spin
variables are defined so that total spin is conserved, but no configuration
space is defined, on which corresponding position variables live. As a
result, the equations of motion cannot be derived through variation of some
action. The second problem is, that the constraint that is intended to
reduce the number of independent spin variables from six to three, fails to
be handled easily. In this short article we show that by taking the missing
spin configuration space to be the Lorentz group, we obtain a variational
principle and a full canonical formalism. We show that the unconstrained
theory is in agreement with the main ideas of the standard model of
electroweak interactions; the six independent variables describing spin as
well as isospin.

Classically spin is described in a Lorentz covariant theory by an
antisymmetic tensor $S_{ab};a,b=0,1,2,3$ \cite{Frenkel,Thomas,BTM,Nyborg}.
 The equations of motion for a
particle with four-velocity $u^\mu $ are supplemented with equations for the
spin variables $S_{ab}$, \cite{BTM} 
\begin{equation}
\label{1} 
\begin{array}{l}
\dot u^\mu =\frac emF^\mu {}_\nu u^\nu \\ 
\dot S_{ab}=\frac{ge}{2m}\left[ F_a{}^cS_{cb}+S_{ac}F^{cd}u_du_b\right]
-S_{ac}\dot u^au_b-\left( a\leftrightarrow b\right) 
\end{array}
\end{equation}
subject to the constraints 
\begin{equation}
\label{2}u^\mu u_\mu =c^2,\quad S_{ab}u^b=0. 
\end{equation}
The total spin is conserved: 
\begin{equation}
\label{4}S_{ab}S^{ab}=const. 
\end{equation}
Allowing for the inclusion of gravitational fields we distinguish between
coordinate (holonomic) indices, denoted by $\mu ,\nu ,...=0,1,2,3\,$ and
Lorentz (anholonomic) indices $a,b,...=0,1,2,3$, which are raised and
lowered by the (possibly nontrivial) space-time metric $g_{\mu \nu }$ and
the Minkowski (flat space-time, background) metric $\eta _{ab}$,
respectively; $g_{\mu \nu }$ and $\eta _{ab}$ are related through the
existence of a tetrad (Vierbein) field $e_a{}^\mu $ by 
\begin{equation}
\label{5}g^{\mu \nu }=\eta ^{ab}e_a{}^\mu e_b{}^v. 
\end{equation}
The tetrad field may also be used to switch from holonomic to anholonomic
indices and vice versa.

The second constraint in (\ref{2}) restricts the number of independent spin
variables to three, since the time-like components $S_{0i},i=1,2,3$ then
vanish in any particle rest frame. The spin equation (\ref{1}) is
constructed to comply with this constraint. But when one seeks to derive
this equation from a Hamiltonian \cite{vanHolten} or Routhian \cite
{YeeBander}, this constraint is not easy to handle. Poisson-Dirac brackets 
\begin{equation}
\label{6}\left\{ S_{ab},S_{cd}\right\} =\eta _{ac}S_{bd}-\eta
_{bc}S_{ad}+\eta _{bd}S_{ac}-\eta _{ad}S_{bc} 
\end{equation}
are postulated, but since these are not compatible with the constraint, 
modifications are needed \cite{vanHolten,YeeBander}. The brackets (\ref{6})
represent the Lie algebra of the Lorentz group, with the idea that the $%
S_{ab}$ correspond to the infinitesimal generators $\Sigma _{ab}$ of Lorentz
transformations, 
\begin{equation}
\label{7}S_{ab}\stackrel{\rm{first quantization}}{\longrightarrow }\Sigma
_{ab}=\frac 12\left( \gamma _a\gamma _b-\gamma _b\gamma _a\right) 
\end{equation}
in the Dirac representation. Our idea is to take the Lorentz group, or
rather $SL(2,{\bf C})$ as its universal covering, as base manifold for
 position
variables, with the spin variables residing in the cotangent space, since
then the correspondence (\ref{7}) is obtained naturally. We were lead to
this idea also from the theory by Barut and Zanghi \cite{BarutZanghi}.
Modelling after the structure of the Dirac equation, they introduced a
classical spinor of four complex variables $ z=\left( z_\alpha \right)
_{\alpha =1,2,3,4},$ in a way so that $z$ and $\bar z=z^{+}\gamma ^0$ become
canonically conjugate variables. In this theory the relations (\ref{2}) and
 (\ref{4}) 
are equivalent to 
\begin{equation}
\label{9}\bar zz =1,\quad \bar z\gamma ^5z=0, 
\end{equation}
which means that if we arrange the four complex numbers $z_\alpha $ 
into a matrix
\begin{equation}
\label{10}G(z)=\left( 
\begin{array}{cc}
z_1+z_3 & z_2-z_4 \\ 
-z_2^{*}-z_4^{*} & z_1^{*}-z_3^{*} 
\end{array}
\right) 
\end{equation}
with determinant 
\begin{equation}
\label{11}\det G(z)=\bar zz - \bar z\gamma ^5z
\end{equation}
then the condition $\det G(z)=1$ for $G(z)$ to be in $SL(2,C)$ is
equivalent to (\ref{9}). 

So we are motivated to describe a classical spinning particle by a
trajectory $\left( x^\mu (\tau ),\Lambda (\tau )\right) $ on the manifold $%
M={\bf R}^4\times SL(2,{\bf C})$, the twofold covering of the Poincar\'e 
group when
equipped with the usual semidirect product. The parameter $\tau $ is assumed
to be invariant under the action of the Poincar\'e group on itself (which
we do not distinguish from $M$), for the action from the left as well as for
the action from the right. In the framework of gravitation theory, where the
space-time part can be equipped with a nontrivial metric, the action from
the left, which is the usual one defined by 
\begin{equation}
\label{12}\left( a,\Lambda _0\right) \circ \left( x,\Lambda \right) =\left(
\Lambda _0x+a,\Lambda _0\Lambda \right) , 
\end{equation}
becomes a composition of a coordinate transformation $x\longmapsto \Lambda
_0x+a$ and a (global) Lorentz gauge transformation $\Lambda \longmapsto
\Lambda _0\Lambda $, which are separately imbedded into the larger groups of
general coordinate transformations and local Lorentz gauge transformations,
respectively \cite{Utiyama}. Dirac four-spinors have no transformation law
under general coordinate transformations, but Einstein's theory of general
relativity can be extended to these objects as a local gauge theory \cite
{Utiyama,TetrodeWeyl}.

Before we define a Hamiltonian we look at the quantized theory first, since
we like to achieve correspondence with the minimally coupled Dirac
equation. In a geometric approach we take all position variables for equal,
so that the quantum states are taken to be functions on $M$. Functions on
the Lorentz group are obtained by taking coordinates $\omega ^{ab}$ in the
Lie algebra and exponentiate these with any representation $\Sigma _{ab}^D$
of the infinitesimal generators to obtain functions 
\begin{equation}
\label{13}\psi _{rs}^D(\omega )=\left[ \exp \left( \omega ^{ab}\Sigma
_{ab}^D\right) \right] _{rs}=D\left( \Lambda \right) _{rs};1\leq r,s\leq
\dim D. 
\end{equation}
Upon quantization the spin momentum variables $S_{ab}$ become first order
differential operators 
\begin{equation}
\label{14}\hat S_{ab}=f_{ab}^{cd}\left( \omega \right) \frac \partial
{\partial \omega ^{cd}} 
\end{equation}
that act on the spin position coordinates $\omega $ of $M$. According to
general theorems \cite{Lie}, the coefficient functions $f_{ab}^{cd}\left(
\omega \right) $ can be choosen so that not only the $\hat S_{ab}$ satisfy
the commution relations of the Lie algebra $sl(2,{\bf C})$, but also act on 
the
column vectors $\psi _{\cdot d}^D,$ $d$ fixed, by multiplication with the
infinitesimal generator in any representation: 
\begin{equation}
\label{15}\hat S_{ab}\psi _{rs}^D(\omega )=\left( \Sigma _{ab}^D\right)
_{rr^{\prime }}\psi _{r^{\prime }s}(\omega ). 
\end{equation}
This realizes the action from the left of the group $SL(2,{\bf C})$ on $M$; 
we can
also realize the action from the right, whence the row vectors in (\ref{13})
are transformed. We further add one dimension as $S^1$ for the quantized
electric charge and arrive at a Kaluza-Klein type theory in eleven
dimensions on the manifold $M^{\prime }={\bf R}^4\times SL(2,{\bf C})\times
 U(1)$
 where the minimally coupled Dirac equation attains the form 
\begin{eqnarray}
i\gamma ^a\left( e_a{}^\mu \partial _\mu -A_a\hat q-\Omega
_a{}^{bc}\hat S_{bc}\right)&& \psi \left( x,\omega ,\phi \right) \nonumber \\
& &=i\gamma
^aD_a\psi =m\psi .\label{16} 
\end{eqnarray}
Here $\phi $ is the coordinate of $S^1$, $\hat q=\partial /\partial \phi $
the quantized charge, so that $D_a$ is
 a local $U(1)\times SL(2,{\bf C})$-gauge
covariant derivative with gauge potentials $A_a$ and $\Omega _a{}^{bc}$.

Since the spin of a particle field is described through its dependence on
the variables $\omega $, for consistency the same must hold for the
interaction fields that carry spin, so the gauge potentials and the tetrad
field must also depend on $\omega $. These transform under the fundamental
representation of the Lorentz group, so are column vectors with 
\begin{equation}
\label{17}\hat S_{ab}B_c(x,\omega )=\left( \Sigma _{ab}^f\right)
_{cc^{\prime }}B_{c^{\prime }}=\eta _{ac}B_b-\eta _{bc}B_a 
\end{equation}
or tensor fields with the corresponding rule for all indices. Since the
fundamental representation has four column vectors, we have a freedom of
choice for the spin-1 fields, which will be interpreted below.

We now define a Hamiltonian for a particle moving on the manifold $M^{\prime}$,
extending the ansatz by van Holten \cite{vanHolten} to gravitational fields,
and supply the missing variational principle. We look at the mass squared
obtained from the Dirac equation (\ref{16}) iterated : 
\begin{eqnarray}
-m^2\psi & =&\gamma ^a\gamma ^bD_aD_b\psi \label{18} \\  
& &\hspace{-.8cm}
=\left[ \frac{\eta ^{ab}}2\left( D_aD_b+D_bD_a\right) +\frac{\Sigma ^{ab}}%
2\left( D_aD_b-D_bD_a\right) \right] \psi . \nonumber
\end{eqnarray}
Here it is important to note that the multiplication with the $\gamma $%
-matrices commutes with the covariant derivatives $D_a$ since these are
scalar operators. We put the symmetrized operator product in (\ref{18}) in
correspondence with the classical momentum variables, reduce the
anticommutator to 
\begin{equation}
\label{20} 
\begin{array}{l}
\left( D_aD_b-D_bD_a\right) = \\ 
\qquad \left( D_be_a{}^\mu \right) \partial _\mu -\left( D_bA_a\right) \hat
q-\left( D_b\Omega _a{}^{cd}\right) \hat S_{cd} - 
\left( a\leftrightarrow b\right)
\end{array}
\end{equation}
and use the freedom to add terms which contribute to the Dirac equation only
as mass constants, to arrive at a Hamiltonian $H$ which is the
squareroot of a nondegenerate
quadratic form in the momentum variables, 
\begin{equation}
\label{21}H^2=H_0+H_1+H_2 
\end{equation}
with 
\begin{eqnarray} 
H_0&=&\eta ^{ab}P_aP_b,\quad P_a=\left( e_a{}^\mu p_\mu -A_aq-\Omega
_a{}^{bc}S_{bc}\right) \nonumber \\ 
H_1&=&\frac 12S^{ab}\left[ \left( D_be_a{}^\mu \right) p_\mu -\left(
D_bA_a\right) q-\left( D_b\Omega _a{}^{cd}\right) S_{cd}\right] \nonumber \\ 
H_2&=&S_{ab}S^{ab}+q^2=2S_{ij}S^{ij}-2S_{0i}S^{0i}+q^2 . \label{22}
\end{eqnarray}
This Hamiltonian is a function $H\left( x,\omega ,\phi ,p,S,q\right) $ on
the cotangent bundle of $M^{\prime }$ where the external fields do not only
depend on the space-time position $x$, but also on the internal coordinates 
$\omega $ and $\phi $, describing their spin and charge. From the action 
\begin{equation}
\label{22b}I=\int d\tau \left( p_\mu \dot x^\mu +S_{ab}\dot \omega
^{ab}+q\dot \phi -H\right) 
\end{equation}
we derive by independent variation of all variables the canonical equations 
\begin{equation}
\label{23} 
\begin{array}{lll}
\dot x=\frac{\partial H}{\partial p}, & \dot \Lambda =\frac{\partial H}{%
\partial S}, & \dot \phi = 
\frac{\partial H}{\partial q} \\ \dot p=-\frac{\partial H}{\partial x}=-\hat
pH, & \dot S=-\hat SH, & \dot q=-\hat qH. 
\end{array}
\end{equation}
Small variations of the position variables are given by a near-identity
Poincar\'e transformation, 
\begin{equation}
\label{24b}\left( x,\Lambda \right) \rightarrow \left( x^\mu +\delta x^\mu
,\exp (\delta \omega ^{ab}\Sigma _{ab}^f)\Lambda \right) 
\end{equation}
under which functions vary as 
\begin{equation}
\label{24c} 
\begin{array}{l}
f\left( x^\mu +\delta x^\mu ,\exp (\delta \omega ^{ab}\Sigma _{ab}^f)\Lambda
\right) \simeq f\left( x^\mu ,\Lambda \right) \\ 
\qquad +\delta x^\nu \partial _\nu f\left( x^\mu ,\Lambda \right) +\delta
\omega ^{ab}\hat S_{ab}f\left( x^\mu ,\Lambda \right) 
\end{array}
\end{equation}
so that the infinitesimal variation gives the action of the infinitesimal
generator on $H$. The law (\ref{17}) hence gives those contributions from
the external fields to the spin equations which were previously generated by
the brackets (\ref{6}). The canonical equations of motion (\ref{23})
conserve the total spin, and lead to conserved sources for the external
fields, but do not satisfy the constraints (\ref{2}). For a discussion of
the first of these constraints we refer to van Holten \cite{vanHolten}, the
second constraint can be done without, as we now show.

The Hamiltonian (\ref{21}) corresponds to the Laplace-Beltrami operator on
the manifold $M^{\prime }$ with the metric having signature $+---,---+++,-$
. Without external fields we have 
\begin{equation}
\label{25}\triangle _{M^{\prime }}=-g^{\mu \nu }\partial _\mu \partial
_v+\eta ^{ac}\eta ^{bd}\hat S_{ab}\hat S_{cd}+\hat q^2,
\end{equation}
so that the mass squared as eigenvalue of this operator depends on spin and
charge. Any wave function $\psi \left( x,\omega ,\phi \right) $ can be
decomposed in the following way. We start with the Fourier decomposition
with respect to the space-time variables, 
\begin{equation}
\label{27}\psi \left( x,\omega ,\phi \right) =\int\limits_{-\infty
}^{+\infty }dm^2\int\limits_{R^3}\frac{d^3p}{\mid p_0\mid }e^{ip_\mu x^\mu
}\hat \psi _{m^2}\left( p_\mu ,\omega ,\phi \right) .
\end{equation}
Next we use the fact that for a given four-vector $p^\mu $ a $SL(2,{\bf C})$
transformation $\Lambda $ can be decomposed as 
\begin{equation}
\label{28}\Lambda =\left( 
\begin{array}{cc}
\lambda  & 0 \\ 
\zeta  & \lambda ^{-1}
\end{array}
\right) U\left( 
\begin{array}{cc}
\lambda ^{\prime } & 0 \\ 
\zeta ^{\prime } & \lambda ^{\prime -1}
\end{array}
\right) ^{-1}
\end{equation}
where $U$ is an element of $SU(2)$, and the matrices $\left( 
\begin{array}{cc}
\lambda  & 0 \\ 
\zeta  & \lambda ^{-1}
\end{array}
\right) $ and $\left( 
\begin{array}{cc}
\lambda ^{\prime } & 0 \\ 
\zeta ^{\prime } & \lambda ^{\prime -1}
\end{array}
\right) $ with $\lambda ,\lambda ^{\prime }$ real, $\zeta ,\zeta ^{\prime }$
complex, are boosts which transform the apex $\left( m,0,0,0\right) $ into $p
$ and $p^{\prime }=\Lambda ^{-1}p$, respectively \cite{Wigner,Rideau,Thaller}%
. This transformation is given by 
\begin{equation}
\label{28b}
\begin{array}{l}
p=\frac m2(\lambda ^2+\lambda ^{-2}+\zeta \zeta ^{*},\lambda ^2-\lambda
^{-2}-\zeta \zeta ^{*}, \\ 
\qquad \qquad \lambda \zeta +\lambda \zeta ^{*},-\lambda \zeta +\lambda
\zeta ^{*}).
\end{array}
\end{equation}
We may replace the six coordinates $\omega ^{ab}$ by three coordinates $u_i$
parametrizing $SU(2)$ and the three coordinates $\lambda ^{\prime },\zeta
^{\prime }$ \cite{Rideau}: 
\begin{equation}
\label{29}\hat \psi _{m^2}\left( p_\mu ,\omega ,\phi \right) =\hat \psi
_{m^2}^{\prime }\left( p_\mu ,u_i,\lambda ^{\prime },\zeta ^{\prime },\phi
\right) .
\end{equation}
Fourier decomposition on the compact manifold $SU(2)$ now gives a discrete
sum over all irreducible representations $D_s$ of $SU(2)$, 
\begin{equation}
\label{30}\hat \psi ^{\prime }\left( p_\mu ,u_i,\lambda ^{\prime },\zeta
^{\prime },\phi \right) =\sum_{s=n/2}\hat \psi _{\alpha \beta }^{\prime
}\left( p_\mu ,s,\lambda ^{\prime },\zeta ^{\prime },\phi \right)
D_s^{\alpha \beta }(u_i).
\end{equation}
The interesting point is, that any irreducible representation occurs with
the multiplicity of its dimension. The columns of these representations
transform under the action of the group from the left as Wigner's induced
representations \cite{Wigner}, with the coordinates $\lambda ^{\prime
},\zeta ^{\prime }$ left invariant. With an inverse Foldy-Wouthysen
transformation these representations can be brought into the standard Dirac
form, for any spin \cite{Mukunda}. Especially
we get spin-1/2 particles in doublets.
 The action from the right transforms the
coordinates $\lambda ^{\prime },\zeta ^{\prime }$ and rotates the
multiplets. If we restrict the action to the $SU(2)$ subgroup of $SL(2,C)$,
the transformations do not depend on the momentum variables, $SU(2)_{\rm{%
right}}$ is a global gauge group. The transformation of the invariant volume
element of the coordinates $p^{\prime },\phi $ under the transformation (\ref
{28b}) into an $SU(2)$-invariant volume element, 
\begin{equation}
\label{31a}\frac{d^3p^{\prime }}{p_0^{\prime }}d\phi =m^2\lambda d\lambda
d\zeta d\phi =d\zeta _1d\zeta _2,
\end{equation}
suggests to introduce two complex coordinates 
\begin{equation}
\label{31}\zeta _1=m\zeta ^{\prime },\quad \zeta _2=m\lambda ^{\prime
}e^{i\phi }
\end{equation}
on which the group $U(1)\times SU(2)_{\rm{right}}$ acts naturally. Thereby
the generator of the electromagnetic gauge group becomes $\left( 1-\tau
_3\right) /2$, as in the standard model of electroweak interactions, and the
spin-1/2 doublet can be interpreted as isospin doublet. So we see that six
independent spin variables naturally lead to a quantized theory describing
spin and isospin. The difficulty to implement the second constraint in (\ref
{2}) can therefore be understood as follows. The overall commuting charge in
this theory is not the electric charge, but the hypercharge, which may
effect isospin rotations, thereby obstructing this constraint.

We obtain the full structure of a local $U(1)\times SU(2)_{\rm{right}}$
gauge theory by replacing the coupling in (\ref{16}) by 
\begin{equation}
\label{32}A_a\hat q\rightarrow gB_a^b\hat \tau _b 
\end{equation}
where $g$ is a coupling constant, the operators $\hat \tau _b=\left( \hat
\tau _0,\hat \tau _{i\,}\right) $ are differential operators in the
coordinates $\zeta _1,\zeta _2$, representing the Lie algebra of $U(1)\times
SU(2)_{\rm{right}}$, acting on any representation as multiplication with
the matrix representation of the infinitesimal generators in analogy to
relation (\ref{15}). The $B_a^b$ are four real local gauge fields, a singlet
and a triplet, taken to be the four column vectors of the fundamental
representation, with $B^1$ and $B^2$ given phase factors $\sin \phi $ and $%
\cos \phi $, respectively, accounting for their electric charge.

In this model the topology breaks the $U(1)\times SU(2)_{\rm{right}}$%
-symmetry in the way of the standard model, since the point $\lambda
^{\prime }=0$ does not correspond to any point in the manifold $M^{\prime }$%
. This reduces finite gauge transformations to the subgroup $U(1)\times U(1)$
generated by $\hat \tau _0$ and $\hat \tau _3$. The different topology with
repect to $\zeta _1$ and $\zeta _2$ gives rise to symmetry breaking effects
through the boundary conditions which we have to impose. We may generate a
discrete mass spectrum by equipping these internal variables with harmonic
oszillator potentials, as was already considered by Dirac in a model with
two real variables \cite{Mukunda,Dirac2}. Then, even with identical
oszillator frequencies, the boundary condition that states have to vanish at 
$\zeta _2=0$, eliminates the uncharged ground state in favour of two massive
charged states. This reproduces also the chiral structure of the standard
model with two neutrinos, for which we have the left- and righthanded
uncharged ground states of the oszillator on $\zeta _1$, and the electron
and the positron, as the two oppositely charged groundstates of the
oszillator on $\zeta _2$ with left-handed and right-handed components each.

Concluding, we have shown that classical spinning particles are naturally
described on an eleven-dimensional manifold with $SL(2,C)\times U(1)$ as
internal space. We proposed a Hamiltonian chosen such that first
quantization leads to a minimally coupled unified Kaluza-Klein type theory
with a local $SL(2,C)_{\rm{left}}\times U(1)\times SU(2)_{\rm{right}}$
gauge group, with a broken $U(1)\times SU(2)_{\rm{right}}$ part that is
nonlinearly related to the $SL(2,C)_{\rm{right }}$ group.

The author thanks Larry P. Horwitz for his hospitality and stimulating
discussions during a stay at Tel Aviv University.

\end{document}